\documentclass[pdftex, a4paper, 12pt]{article}
\usepackage{algos}
\usepackage[square, numbers]{natbib}

\newcommand{\mS}{\mathcal{S}}

\newcommand{\mA}{\mathcal{A}}

\newcommand{\mH}{\mathcal{H}}

\title{Optimally Installing Strict Equilibria}

\author{Jeremy McMahan\thanks{University of Wisconsin-Madison. Corresponding author: \href{mailto:jmcmahan@wisc.edu}{\texttt{jmcmahan@wisc.edu}}}, Young Wu\footnotemark[1], Yudong Chen\footnotemark[1], Xiaojin Zhu\footnotemark[1], \\ and Qiaomin Xie\footnotemark[1]}

\begin{document}

\maketitle

\begin{abstract}

    In this work, we develop a reward design framework for installing a desired behavior as a strict equilibrium across standard solution concepts: dominant strategy equilibrium, Nash equilibrium, correlated equilibrium, and coarse correlated equilibrium. We also extend our framework to capture the Markov-perfect equivalents of each solution concept. Central to our framework is a comprehensive mathematical characterization of strictly installable, based on the desired solution concept and the behavior's structure. These characterizations lead to efficient iterative algorithms, which we generalize to handle optimization objectives through linear programming. Finally, we explore how our results generalize to bounded rational agents.
\end{abstract}

\section{Introduction}\label{sec: intro}

Mechanism Design has been critical for remarkable achievements in online advertising, fair resource allocation, traffic routing, and many other applications. However, many classical mechanisms exhibit a crucial security flaw: the desired behavior of the agents is only weakly dominating. Consequently, bounded rational agents would be willing to deviate, leading to unpredictable behaviors. These effects can be especially devastating in sequential decision-making settings since the impact of rogue behavior can compound over time. To address these issues, we study the design of sequential mechanisms that strictly enforce a desired behavior. 

We capture these issues broadly through the lens of \emph{optimal reward design}. Formally, suppose a game designer wishes to construct a Markov game $G$ for which a desired behavior $\pi$ is a strict solution under a given solution concept, such as dominant strategy equilibrium (DSE), Nash equilibrium (NE), correlated equilibrium (CE), or course correlated equilibrium (CCE). However, the game designer can only choose the reward function, which must obey a reward bound. This formulation models realistic scenarios where the structure of the environment is hard to change, but the rewards are easy to change, such as in the example of safe traffic control through tolls rather than a road network overhaul. Moreover, if the designer has some objective function, it must choose a feasible reward function while optimizing this objective.

It is known that the only behaviors that can be a strict NE in games are pure strategies. However, if we restrict the types of deviations possible, more behaviors can be strictly enforced. Moreover, the more complex, recommendation-based solution concepts of the CE and CCE allow for even more complicated behaviors to be strictly enforced and with a larger margin of strictness. Understanding these complexities alone is challenging, and this is further exacerbated by adding an optimization objective and considering Markov games.

\paragraph{Our Contributions.} We present the first-of-its-kind complete characterizations of behaviors that are strictly enforceable for CE and CCE. We also analyze the maximum strictness gap possible as a function of the reward bound, the structure of the target behavior, and the structure of the deviation behavior. In addition, we present sufficient conditions for a policy to be a strict solution for each solution concept's Markov-perfect equivalent. Then, we leverage our insights into the design of near-linear time algorithms for determining the strict enforceability of a given behavior and producing a simple feasible solution, should one exist. Lastly, we develop polynomial-sized linear programs to solve the optimal reward design problem.

\subsection{Related Work}
\label{sec: related}

Standard mechanism design ~\cite{tadelis2013game, roughgarden2010algorithmic, osborne1994course} focuses on the implementation of a weak dominant strategy equilibrium or a Bayesian Nash equilibrium, both of which allow weak preference of a player's equilibrium strategy over another action. This implies that in equilibrium, the players could potentially deviate to another strategy without changing their expected payoffs, which is undesirable.

Reward design for single-agent Markov decision process has been studied in \citet{banihashem2022admissible, huang2019deceptive, rakhsha2021policy, rakhsha2021reward, rakhsha2020policy, zhang2020adaptive}, and the data poisoning problem in this setting has also been studied in~\citet{ma2019policy, rangiunderstanding, zhang2008value, zhang2009policy}. When there is only one agent, a deterministic optimal policy always exists, and it consists of actions that are weakly preferred to all other actions in every state and every period. As a result, these techniques do not extend to the reward design or data poisoning for Markov games.

Recent work on data poisoning for multi-agent Markov games considers the installation of a strict dominant strategy equilibrium ~\cite{wu2023reward, ma2021game} or a strict Nash equilibrium ~\cite{wu2024data} as the unique equilibrium of some Markov game. These papers study the problem of modifying offline training datasets instead of directly changing the rewards. ~\cite{monderer2003k, anderson2010internal} study the problem of installing a pure strategy equilibrium by adjusting the payoff matrix while minimizing the modification. Still, their method does not directly extend to solution concepts like correlated equilibria and coarse correlated equilibria. Another recent work ~\cite{wu2023minimally} studies the problem of installing a unique Nash equilibrium, possibly in mixed strategies. However, their results only apply to zero-sum Markov games, whereas our paper also applies to general-sum games.

\section{Strict Equilibria}\label{sec: strictness}

\paragraph{Markov Games.} A (tabular, finite-horizon) $n$-player \emph{Markov Game} (MG) is a tuple $G = (\mS, \mA, P, r, H)$, where (i) $\mS$ is the finite set of \emph{states}, (ii) $\mA = \mA_1 \times \cdots \mA_n$ is the finite set of \emph{joint actions}, (iii) $P_h(s, a) \in \Delta(S)$ is the \emph{transition} distribution at time $h \in [H]$, (iv) $r_h(s, a) \in \Real^n$ is the \emph{reward} function at time $h\in[H]$, and (v) $H$ is the finite time \emph{horizon}. 

\paragraph{Interaction Protocol.} 
The agents interact with $G$ using a \emph{joint policy} $\pi = \{\pi_h\}_{h = 1}^H$. In general, $\pi_{h}: \mH_h \to \Delta(\mA)$ is a mapping from the observed history at time $h$ to a distribution of actions. Often, researchers study \emph{Markovian policies}, which take the form $\pi_h : \mS \to \Delta(\mA)$, and \emph{product policies}, which take the form $\pi = \{\pi_i\}_{i =1}^n$, where each $\pi_i$ is an independent policy for the $i$th player. 

The players start in an initial state $s_1 \in \mS$ with an observed history $\tau_1 = (s_1)$. For any $h \in [H]$, the players choose a joint action $a_h \sim \pi_h(\tau_h)$. The players then receive an immediate reward vector $r_h \sim R_h(s,a)$. Lastly, $G$ transitions to state $s_{h+1} \sim P_h(s_h,a_h)$, prompting the players to update their observed history to $\tau_{h+1} = (\tau_h, a_h, s_{h+1})$. This process is repeated for $H$ steps; the interaction ends once $s_{H+1}$ is reached.

\paragraph{Solution Concepts.} Solutions to games take the form of \emph{equilibrium}. An equilibrium concept ensures that players do not benefit from changing their policy, assuming the policies of the other players. This intuition leads to the concepts of \emph{Nash equilibrium} (NE), \emph{correlated equilibrium} (CE), and \emph{coarse-correlated equilibrium} (CCE). In the Multi-Agent Reinforcement Learning (MARL) realm, most work focuses on the Markov-perfect variations of each concept. In this work, we further focus on strict variations of each equilibrium.

\begin{definition}[Strict Equilibria]
    We call a Markovian policy $\pi$ a \emph{strict Markov-perfect course correlated equilibrium} (sMPCCE) for an MG $G$ if for all players $i \in [n]$, times $h \in [H]$, states $s \in \mS$, and deviation policies $\pi_i'$, 
    \begin{equation}\label{equ: smpe}\tag{sMPCCE}
        V_{i,h}^{\pi}(s) > V_{i,h}^{\pi_i', \pi_{-i}}(s),
    \end{equation}
    where, $V^{\pi}_{i,h}(s) \defeq \E^{\pi}_G\brak{\sum_{t = h}^H r_{i,t} \mid s_h = s}$ denotes the value of $i$ from interacting with $G$ using $\pi$, and $\E^{\pi}_G$ denotes the expectation defined by the probability law over full histories, $\P^{\pi}_G$. 
    Furthermore, we call $\pi$ an \emph{strict Markov-perfect (Nash) equilibrium} (sMPE or sMPNE) for $G$ if $\pi$ is additionally a product policy (over the $n$ players). 
    
    We call $\pi$ an \emph{strict  Markov-perfect correlated equilibrium} (sMPCE) for an MG $G$ if for all players $i \in [n]$, times $h \in [H]$, states $s \in \mS$, and deterministic strategy modifications $\phi = \{ \phi_{h,s} : \mA_i \to \mA_i\}_{h,s}$, 
    \begin{equation}\label{equ: smpce}\tag{sMPCE}
        V_{i,h}^{\pi}(s) > V_{i,h}^{\phi \circ \pi}(s).
    \end{equation}
    Here, $\phi \circ \pi$ denotes the policy induced at each stage $(h,s)$ by drawing an action $a \sim \pi_{h}(s)$ and then using the action $(\phi_{h,s}(a_i), a_{-i})$.
\end{definition}

\paragraph{The Power of Strictness.} We focus on strict equilibria since they ensure predictable outcomes. This is because any rational agent would be unwilling to deviate if it means suffering worse payoffs. Thus, a game designer can guarantee any desired behavior $\pi$ so long as the agents know $\pi$ is a strict equilibrium. The designer can orchestrate this scenario by explicitly recommending that agents play $\pi$ when releasing the constructed game to the public, as standard in many mechanism design works. The agents could then efficiently check $\pi$ is indeed a strict equilibrium for the game. After verification, the only rational choice for the agents would be to follow $\pi$.

\begin{observation}[Predictable Outcomes]\label{obs: outcomes}
    Suppose that $G$ is any Markov game and $\pi$ is an sMPCCE for $G$. Then for any player $i \in [n]$, if $i$ is rational and believes that all other players play according to $\pi$, then player $i$ will \emph{uniquely} play according to $\pi$.
\end{observation}

\begin{remark}[Strictness Trade-off]
    Strictness is the ultimate goal for a game designer. However, there is a good reason why most works do not design strict equilibria: doing so is generally impossible. As we will see later, only pure strategies can be a strict NE, drastically reducing the pool of possible behaviors. The key development of our work is relaxing the solution concept or the strictness constant, which can enable much larger classes of behavior to be installed. Thus, strictness is a reasonable final goal given our new insights and the ability of a designer to recommend a desired behavior.
\end{remark}

\paragraph{Reward Design.} In many environments, transitions are much harder to change than rewards. Imagine a government trying to change traffic flow through a road network. It would be much cheaper to modify tolls on existing roads to manipulate traffic than to change the road network. Consequently, in this work, we suppose the game designer can only choose the rewards for an already known transition structure. 

\begin{definition}[Strict Installability]
    For any solution concept SOL, $\pi$ is \emph{SOL-strictly installable} if there exists a game for which $\pi$ is a strict SOL.
\end{definition}

Here, we use the term "installable" for Markov games without player types, instead of "implementable" from mechanism design literature, to avoid confusion.

\begin{definition}[Optimal Reward Design]\label{def: reward-design}
    In the \emph{optimal reward design} problem, we are given a desired behavior policy $\pi$, a reward bound $B$, a reward-less Markov game $G$, and a desired solution concept $\text{SOL} \in \{\text{MPE}, \text{MPCE}, \text{MPCCE}\}$. The designer's goal is to compute a reward function $r$ so that $\pi^{\dagger}$ is a strict-SOL for $G[r]$, the Markov game induced by augmenting $G$ with rewards $r$. Moreover, the designer wishes to minimize some cost function associated with the new game, $C^{\pi}(r)$. Overall, we formulate the optimal reward design problem as,
    \begin{equation}\tag{ORD}\label{equ: ord}
    \begin{split}
        \min_{r \in \Real^{H\times S \times A}} \quad & C^{\pi}(r) \\
        \text{s.t.}  \quad & \text{$\pi$ is a strict-SOL for $G[r]$} \\
        &r_h(s,a) \in [-B, B] \quad \quad \forall h,s,a
    \end{split}
    \end{equation}
\end{definition}

\begin{example}[Costs]\label{ex: costs}
We consider two natural cost settings in this paper. In the first setting, which is common in adversarial reinforcement learning, there is an initial reward function $r^{0}$ that the designer is modifying. In the second setting, common in game theory, there is no initial reward function to consider. Each setting gives rise to different cost considerations.
\begin{enumerate}
    \item Suppose $G$ had an initial reward function $r^{0}$. Then, it is often desirable to design the new reward function to be as close to $r^{0}$ as possible. We consider two fundamental measures of closeness.

    \begin{enumerate}
        \item \emph{Online Cost.}
        \begin{equation}
            C^{\pi}(r) = \sum_{h,s,a} \mu_h(s,a) \abs{r_h(s,a) - r^{0}_h(s,a)},
        \end{equation}
        Here, $\mu_h$ corresponds to the visitation measure induced by the policy $\pi.$
        For example, a government changing known traffic rules would be incentivized to minimize the amount of change to promote a smooth transition to the new rules. 
        \item \emph{Offline Cost.}
        \begin{equation}
            C^{\pi}(r) = \sum_{h,s,a} \abs{r_h(s,a) - r^{0}_h(s,a)}
        \end{equation}
        Similarly, in an adversarial setting, an attacker would want to minimize the amount of data corruption to avoid triggering detection. 
    \end{enumerate}

    \item If there is no initial reward, various social welfare functions could be considered so that players feel $\pi$ is also fair.
    \begin{enumerate}
        \item \emph{Social welfare maximization.}
        \begin{equation}
            C^{\pi}(r) = -\sum_{i} V_{G[r], i}^{\pi}.
        \end{equation}
        This installs the desired target in the best way for the collective. 
        \item \emph{Egalitarian welfare maximization. }
        \begin{equation}
            C^{\pi}(r) = -\min_{i} V_{G[r], i}^{\pi}.
        \end{equation}
        Ensures the worst-off player is not too bad off.
    \end{enumerate}
\end{enumerate}
\end{example}

\section{Feasibility Characterizations}\label{sec: feasibility}

In this section, we explore the behavior profiles that can be installed for each solution concept. To this end, we exactly characterize strictly installable strategies for normal-form games. We proceed in the order of most restrictive characterizations to least restrictive. We then use these characterizations to derive sufficient conditions for strongly installable strategies. Lastly, we extend these results to Markov games.  

To build intuition, we start with the simple normal-form game setting. We represent a $n$-player general-sum game by a pair $(\mA, u)$. Here, $\mA \defeq \mA_1 \times \cdots \times \mA_n$ denotes the finite joint action space for the players, and $u_i(a)$ denotes the utility of player $i$ from joint action $a$. For any player $i \in [n]$, action $a_i \in \mA_i$, and mixed strategy $\sigma \in \Delta(\mA)$, we say that $a_i$ is $\sigma$-supported if there exists some $a_{-i} \in \mA_{-i}$ such that $\sigma(a_i, a_{-i}) > 0$. The marginal distributions of the actions will play a vital role in our characterizations.

\begin{definition}[Conditionals]\label{def: conditional}
    For a given mixed strategy $\sigma \in \Delta(\mA)$, player $i \in [n]$, and action $j \in \mA_i$, we let $p_{ij} \defeq \sum_{a_{-i} \in \mA_{-i}} \sigma(j, a_{-i})$ denote the probability that player $i$ plays $j$ under $\sigma$. Then, we refer to the \emph{conditional} of $\sigma$ by the mixed strategy over $\mA_{-i}$ defined by,
    \begin{equation}
        \sigma_{ij}(a_{-i}) \defeq \sigma(j, a_{-i})/p_{ij}.
    \end{equation}
    If $p_{ij} = 0$, we define $\sigma_{ij} \defeq 0$ to be the all-zero vector.
\end{definition}

\subsection{sNE}\label{subsec: sne}
We begin by discussing the arguably most famous solution concept: the Nash Equilibrium (NE). Since NE exhibits the strongest requirements of the three equilibria, it naturally allows the fewest strategies to be strictly installable. In fact, it is well known that only pure strategies can be strictly installed. For completeness and to motivate future solution concepts, we rewrite this condition in terms of the distributional structure of $\sigma$. 

\begin{proposition}[sNE installability]\label{prop: sne}
    Given any mixed product strategy $\sigma \in \Delta(\mA)$, there exists a utility function $u$ for which $\sigma$ is an sNE for $(\mA, u)$ if and only if for all players $i \in [n]$ we have that $p_{ij} = 1$ for some $j \in \mA_i$. 
\end{proposition}

\subsection{sCE}\label{subsec: sce}

We next move on to sCE. The key observation is that the strictness constraints for any two supported actions $j$ and $k$ are in opposition.
Specifically, for each player $i$, the strictness constraint when recommended action $j$ over deviating to action $k$ requires the following:
\begin{equation}
    \sum_{a_{-i} \in \mA_{-i}} \sigma_{ij}(a_{-i}) \paren{u_{i}(j,a_{-i}) - u_i(k,a_{-i})} > 0,
\end{equation}
which states that the utility for action $j$ is generally better than for action $k$. Similarly, the strictness constraint when recommended $k$ over deviating to $j$ requires
\begin{equation}
    \sum_{a_{-i} \in \mA_{-i}} \sigma_{ik}(a_{-i}) \paren{u_{i}(j,a_{-i}) - u_i(k,a_{-i})} < 0.
\end{equation}
If $\sigma_{ij} = \sigma_{ik}$, then the summation term must be positive and negative simultaneously, a contradiction. This shows that differing conditionals on supported actions is a necessary condition.

We can also show that differing conditionals are sufficient. The assumption that the conditionals are different can be directly exploited by defining the utility to match the normalized conditional: $u_i(j, a_{-i}) \defeq \sigma_{ij}(a_{-i})/\norm{\sigma_{ij}}_2$. This construction ensures the expected utility when deviating to $k$ is roughly $1 - \cos(\theta_{ijk})$, where $\theta_{ijk}$ is the angle between $\sigma_{ij}$ and $\sigma_{ik}$. This quantity is uniquely maximized when $\sigma_{ik} = \sigma_{ij}$. Since the conditionals are assumed to be distinct, the optimal solution is thus $\sigma_{ij}$. Consequently, player $i$ has a strict incentive to follow the recommendation.

\begin{theorem}[sCE installability]\label{thm: sce}
    Given any mixed strategy $\sigma \in \Delta(\mA)$, there exists a utility function $u$ for which $\sigma$ is an sCE for $(\mA, u)$ if and only if for all players $i \in [n]$ and all pairs of $\sigma$-supported actions $j,k \in \mA_i$ $(j\neq k)$, $\sigma$ satisfies $\sigma_{ij} \neq \sigma_{ik}$. 
\end{theorem}

\begin{proof}

    $[\implies]$ We prove the contrapositive. Suppose that there exists a player $i$, $\sigma$-supported action $j$, and deviation action $k$, satisfying $\sigma_{ij} = \sigma_{ik}$. For any $a_{-i}$, let $d(a_{-i}) \defeq u_{i}(j,a_{-i}) - u_i(k,a_{-i})$. We observe that the sCE condition for $i$ to $k$ translates to,
    \begin{equation}
        \sum_{a_{-i} \in \mA_{-i}} \sigma_{ij}(a_{-i}) d(a_{-i}) > 0.
    \end{equation}
    Moreover, the sCE condition for $k$ to $i$ translates to,
    \begin{equation}
        \sum_{a_{-i} \in \mA_{-i}} \sigma_{ik}(a_{-i}) d(a_{-i}) < 0.
    \end{equation}
    Now, using the fact that $\sigma_{ij} = \sigma_{ik}$, we observe that,
    \begin{equation}
        0 < \sum_{a_{-i} \in \mA_{-i}} \sigma_{ij}(a_{-i}) d(a_{-i}) = 
        \sum_{a_{-i} \in \mA_{-i}} \sigma_{ik}(a_{-i}) d(a_{-i}) < 0.
    \end{equation}
    This completes the contrapositive.

    $[\impliedby]$ Suppose that for all players $i \in [n]$, all $\sigma$-supported actions $j \in \mA_i$, and all deviation actions $j \neq k \in \mA_i$, $\sigma$ satisfies $\sigma_{ij} \neq \sigma_{ik}$. 
    We explicitly construct a utility function $u$ by $u_i(j, a_{-i}) \defeq \frac{\sigma_{ij}(a_{-i})}{\norm{\sigma_{ij}}_2}$. Also, for any deviation action $k$, let $\theta_{ijk}$ denote the angle between the two vectors $\sigma_{ij}$ and $\sigma_{ik}$. Then, we see that, 
    \begin{align*}
        \sum_{a_{-i} \in \mA_{-i}} \sigma_{ij}(a_{-i}) d(a_{-i}) &= \sum_{a_{-i} \in \mA_{-i}} \sigma_{ij}(a_{-i})\Big( \frac{\sigma_{ij}(a_{-i})}{\norm{\sigma_{ij}}_2}-\frac{\sigma_{ik}(a_{-i})}{\norm{\sigma_{ik}}_2}\Big) \\
        &= \norm{\sigma_{ij}}_2 - \frac{ \inner{\sigma_{ij}, \sigma_{ik}}}{\norm{\sigma_{ik}}_2} \\
        &= \norm{\sigma_{ij}}_2 - \norm{\sigma_{ij}}_2\cos(\theta_{ijk})\\
        &= \norm{\sigma_{ij}}_2(1 - \cos(\theta_{ijk})) \\
        &> 0.
    \end{align*}
    The inequality follows since $\cos(\theta_{ijk}) < 1$. Specifically, since all vectors are in the positive orthant and $\sigma_{ij} \neq \sigma_{ik}$, we have that $\theta_{ijk} > 0$ and $\theta_{ijk} < \pi$, both together imply that $\cos(\theta_{ijk}) < 1$.
    
\end{proof}

\paragraph{Algorithmic Interpretation.} Notice that our mathematical characterization for sCE translates into an iterative algorithm for determining installability. We can verify if $\sigma$ is sCE-installable by looping through all pairs of actions and checking if their conditionals are the same. Should we determine that $\sigma$ is sCE-installable, we can also produce a witness utility function efficiently through our explicit construction. 

\begin{corollary}\label{cor: sce}
    Determining if a given $\sigma \in \Delta(\mA)$ is sCE-installable can be performed in $O(n\max_i |\mA_{-i}||\mA_i|^2)$ time, which is polynomial in the input size. Moreover, if $\sigma$ is sCE-installable, then the simple utility function $u_i(j, a_{-i}) \defeq \sigma_{ij}(a_{-i})/\norm{\sigma_{ij}}_2$ witnesses $\sigma$'s sCE-installability. 
\end{corollary}

\subsection{sCCE}\label{subsec: scce}

Unlike sCE, having some equal conditionals does not necessarily preclude strictness. This follows since the sCCE condition considers deviations before a recommendation is instantiated. Specifically, the strictness constraint for deviation $k$ is,
\begin{equation}
    \sum_{\ell\in \mA_i}\sum_{a_{-i} \in \mA_{-i}} \sigma(\ell, a_{-i}) (u_i(\ell, a_{-i}) - u_i(k, a_{-i})) > 0.
\end{equation}
Importantly, even if some action $j$ were dominated in utility by $k$, the other supported actions $\ell$ can help to ensure $k$ is dominated overall. Thus, it is much harder for a deviation to be strictly preferable, allowing more diverse behavior profiles to be installed. 

In fact, we show that sCCE only requires \emph{one} pair of supported actions to differ. The key idea is that we can use both actions in conjunction to dominate any other deviation while carefully balancing the utility so that none is large enough to be a viable deviation alone. Using the same definition of utility as before, the expected utility difference for deviation $m$ can be roughly lower bounded by,
\begin{equation}
    p_{ij}(1 - \cos(\theta_{ijm})) + p_{ik}(1 - \cos(\theta_{ikm})).
\end{equation}
Critically, one of these terms must be non-zero since $\sigma_{ij} \neq \sigma_{ik}$ by assumption. This implies the sufficiency of the condition. The necessity follows as before for sCE since if all conditionals are equal, we would again have contradictory inequality demands.

\begin{theorem}[sCCE installability]\label{thm: scce}
    Given any mixed strategy $\sigma \in \Delta(\mA)$, there exists a utility function $u$ for which $\sigma$ is an sCCE $(\mA, u)$ if for all players $i \in [n]$, either $\sigma$ supports only one action in $\mA_i$ or there exist two $\sigma$-supported actions $j \neq k \in \mA_i$ satisfying $\sigma_{ij} \neq \sigma_{ik}$.
\end{theorem}

\begin{proof}

    Suppose that for all players $i \in [n]$, there exist two actions $j \neq k \in \mA_i$ satisfying $\sigma_{ij} \neq \sigma_{ik}$. At least one of these actions must be supported by $\sigma$ for the marginals to differ, and WLOG, we assume $j$ is $\sigma$-supported. We explicitly construct a utility function $u$ by $u_i(j, a_{-i}) \defeq \frac{\sigma_{ij}(a_{-i})}{\norm{\sigma_{ij}}_2}$. Then, for any deviation action $m$, we see that,
    \begin{align*}
        \sum_{\ell\in \mA_i}\sum_{a_{-i} \in \mA_{-i}} \sigma(\ell,a_{-i}) d(a_{-i}) &= \sum_{\ell\in \mA_i}\sum_{a_{-i} \in \mA_{-i}} \sigma(\ell, a_{-i}) \Big(\frac{\sigma_{i\ell}(a_{-i})}{\norm{\sigma_{i\ell}}_2} - \frac{\sigma_{im}(a_{-i})}{\norm{\sigma_{im}}_2}\Big) \\
        &= \sum_{\ell \in \mA_i} p_{i\ell} \Big(\norm{\sigma_{i\ell}}_2 - \frac{\inner{\sigma_{i\ell}, \sigma_{im}}}{\norm{\sigma_{im}}_2}\Big) \\
        &= \sum_{\ell \in \mA_i} p_{i\ell} \norm{\sigma_{i\ell}}_2 \Big(1 - \cos{(\theta_{i\ell m})}\Big) \\
        &\geq p_{ij}\norm{\sigma_{ij}}_2 \Big(1 - \cos{(\theta_{ij m})}\Big) \\
        &+ p_{ik}\norm{\sigma_{ik}}_2 \Big(1 - \cos{(\theta_{ikm})}\Big) \\
        &> 0.
    \end{align*}
    The equalities follow from using the same argument for sCE but doing this for each $\ell$. For the strict inequality, first observe that by the support assumption $p_{ij}, p_{ik} > 0$, and so $\norm{\sigma_{ij}}_2, \norm{\sigma_{ik}}_2 > 0$. Thus, the only way for the inequality to fail would be for both $1 - \cos(\theta_{ijm}) = 0 = 1 - \cos(\theta_{ijk})$. By
    the sCE argument, a strict inequality holds if $\sigma_{im} \neq \sigma_{ij}$. On the other hand, if $\sigma_{im} = \sigma_{ij}$, the first term is $0$ but the second term must be $> 0$ since by assumption $\sigma_{im} = \sigma_{ij} \neq \sigma_{ik}$.

    If instead only one action $j$ is supported, it is easy to see that using the same utility construction above, that $u_i(k, a_{-i}) = 0$ always holds, whereas at least one $u_i(j, a_{-i}') > 0$. Thus, strict dominance holds in all cases.
\end{proof}

\paragraph{Algorithmic Interpretation.} Again, our mathematical characterization translates into an iterative algorithm. In the case of sCE, finding a single pair of non-zero conditionals was sufficient to rule out installability. Here, we need to guarantee that \emph{all} non-zero conditionals are equal to rule out installability. Fortunately, this can be performed even faster than before, in linear time, with simple for loops. Moreover, the same utility we used as a witness for sCE works just as well for sCCE.

\begin{corollary}\label{cor: scce}
    Determining if a given $\sigma \in \Delta(\mA)$ is sCCE-installable can be performed in $O(n|\mA|)$ time, which is linear in the input size. Moreover, if $\sigma$ is sCCE-installable, then the simple utility function $u_i(j, a_{-i}) \defeq \sigma_{ij}(a_{-i})/\norm{\sigma_{ij}}_2$ witnesses $\sigma$'s sCCE-installability.
\end{corollary}

\subsection{sMPCCE}\label{subsec: smpe}

We now extend all the results we have seen so far to the Markov game setting. As before, our results will depend heavily on the structure of the conditionals. However, we must now consider the conditionals of each stage game.

We show in general that strictly installing Markov-perfect equilibria boils down to strictly installing the partial policy in each stage game. Thus, we can directly apply our previous results per stage. 

\begin{theorem}[sMPCCE installability]\label{thm: sMPCCE}
    Given any policy $\pi \in \Pi$, there exists a reward function $r$ for which $\pi$ is a strict Markov-perfect SOL (NE/CE/CCE) for $G[r]$ if for all stages $(h, s)$, the mixed strategy $\pi_h(s)$ satisfies the conditions for strict SOL (NE/CE/CCE).
\end{theorem}

\begin{corollary}\label{cor: sMPCCE}
    Sufficient sMPCCE-installability can be checked in polynomial time by running the corresponding algorithm for the desired solution concept at each stage game. Moreover, if $\pi$ is sMPCCE-installable, a feasible reward function can be constructed by pairing the feasible utilities from before at each stage game. 
\end{corollary}

\section{Efficient Optimization}\label{sec: optimization}
In this section, we design polynomial-time algorithms for \eqref{equ: ord}. First, we present linear program formulations for the normal-form game case. Then, we combine these linear programs with a backward induction idea to solve the Markov game case.

Linear programming can solve classical solution concepts since their defining constraints are linear inequalities. However, the strict inequalities may induce a non-polytope feasible set. Fortunately, as we showed in \cref{sec: feasibility}, strict equilibrium can be captured with a non-strict linear inequality with a sufficiently small slack variable, $\iota$. For example, we can solve \eqref{equ: ord} for any linear objective and solution concept sCCE with the following LP:
\begin{equation}\label{equ: sccelp}
    \begin{aligned}
        \min_{u} \quad & C^{\pi}(u) \\
        \text{s.t.} \quad & \sum_{a \in \mA} \sigma(a) \left(u_i(a) - u_i(a_i', a_{-i}) \right) \geq \iota \\
        & -B \leq u_i(a) \leq B
    \end{aligned}
\end{equation}
The first constraint is for $\forall i \in [n], a_i' \in \mA_i$, and the second is for $\forall i \in [n], a \in \mA$.
Importantly, $\sigma$ here is fixed, whereas $u$ is the optimizing variable. This ensures the inequalities above are indeed linear. Overall, we then see that reward design can be solved efficiently. 

\begin{proposition}[Normal-form Reward Design]\label{prop: matrix-lp}
    For any normal-form game skeleton $\mA$, mixed strategy $\sigma \in \Delta(\mA)$, slack parameter $\iota > 0$, and reward bound $B$, \eqref{equ: sccelp} is equivalent to \eqref{equ: ord} for sCCE. Thus, if \textsc{LPSolve} is a polynomial-time linear-program solver, then \textsc{LPSolve}\eqref{equ: sccelp} solves \eqref{equ: ord} in polynomial time.
\end{proposition}

\begin{remark}[Extensions]
    A similar LP to \eqref{equ: sccelp} can also be straightforwardly derived for sNE and sCE. 
\end{remark}

As standard for MARL, it is tempting to recursively solve each matrix stage game to solve a Markov game. However, doing so would be suboptimal and may even incorrectly determine that no solution exists. The main issue is that a standard backward induction approach fixes future-stage game designs and has to work around them. In contrast, an optimal solution could leverage future-stage reward designs to enforce strictness at an earlier stage. Thus, it is critical to consider all stages simultaneously.

Although simultaneously considering all stages sounds like an impossible task, we show it can be done with one LP. Critically, since our target $\pi$ is fixed, the equalities defining the $Q$ function are linear in the immediate reward, which is our optimizing variable. Thus, we can use a similar LP as in the matrix game case, but that explicitly operates on the $Q$ functions. The LP is as follows: 

\begin{equation}\label{equ: lp}
    \begin{aligned}
        \min_{u} \quad & C^{\pi}(r) \\
        \text{s.t.} \quad & \sum_{a \in \mA} \pi_h(s, a) \left(Q_{i,h}^{\pi}(s,a) - Q_{i,h}^{\pi}(s, (a_i', a_{-i})) \right) \geq \iota \\
        &Q_{i,h}^{\pi}(s,a) = r_{i,h}(s,a) + \sum_{s' \in \mS} P_h(s' \mid s,a) V_{i,h+1}^{\pi}(s')\\
        &V_{i,h}^{\pi}(s) = \sum_{a \in \mA} \pi_{i,h}(s,a) Q_{i,h}^{\pi}(s,a) \\
        &V_{i,H+1}^{\pi}(s) = 0 \\
        & -B \leq r_{i,h}(s,a) \leq B
    \end{aligned}
\end{equation}
The first constraint is for $\forall i \in [n], h \in [H], s \in \mS, a_i' \in \mA_i$.

\begin{proposition}[MG Reward Design]\label{prop: mg-lp}
    For any Markov-game skeleton $G$, policy $\pi \in \Pi$, slack parameter $\iota > 0$, and reward bound $B$, \eqref{equ: lp} is equivalent to \eqref{equ: ord} for sMPCCE. Thus, if \textsc{LPSolve} is a polynomial-time linear-program solver, then \textsc{LPSolve}\eqref{equ: sccelp} solves \eqref{equ: ord} in polynomial time for Markov Games.
\end{proposition}

\begin{corollary}
    Any of the objectives from \cref{ex: costs} may be used in the above LP as they all are linear for fixed target $\pi$.
\end{corollary}

\section{Strict(er) Installation}

Up until now, we have only considered rational agents. However, in practice, agents are often not rational. If we instead assume agents have bounded rationality, we must boost the strictness gaps to ensure agents adopt the desired behaviors. 

\begin{definition}[Bounded Rationality]\label{def: bounded-rational}
    Player $i$ is $\epsilon$-rational, if given other players behavior profile $\pi_{-i}$, player $i$ is willing to play any $\pi'_i$ that is an $\epsilon$-approximate best response: $V^{\pi_i', \pi_{-i}} \geq \max_{\pi_i} V^{\pi_i, \pi_{-i}} - \epsilon$.
\end{definition}

Given $\epsilon$-bounded rational players, we can still guarantee players play $\pi$ similar to \cref{obs: outcomes} by installing an $\epsilon$-strict equilibrium. 

\begin{definition}[$\epsilon$-Strict Equilibria]
    For any $\epsilon > 0$, we strengthen each strict equilibrium to an $\epsilon$-strict equilibrium, which requires an $\epsilon$-dominance gap. For sMPCCE and sMPNE, we replace \eqref{equ: smpe} with the new constraint:
    \begin{equation}\label{equ: ssmpe}\tag{$\epsilon$-sMPCCE}
        V_{i,h}^{\pi}(s) > V_{i,h}^{\pi_i', \pi_{-i}}(s) + \epsilon.
    \end{equation}
    For sMPCE, we replace \eqref{equ: smpce} with the new constraint:
    \begin{equation}\label{equ: ssmpce}\tag{$\epsilon$-sMPCE}
        V_{i,h}^{\pi}(s) > V_{i,h}^{\phi \circ \pi}(s) + \epsilon.
    \end{equation}
    
\end{definition}

\paragraph{Deviation Classes.} Unlike standard strict equilibria, enforcement $\epsilon$-strictness depends crucially on the players' class of possible deviations. For example, if we wish to install $a^*$ as an $\epsilon$-sNE, this will be impossible if player $i$ plays $a^*_i$ with high probability. In fact, for any given utility function, player $i$ could add more probability mass to $a_i^*$ to break the strictness constraint. 

Consequently, to guarantee $\epsilon$-strictness, we must assume players' deviations never place mass on $a^*$. Similarly, for $\epsilon$-sCE, we must assume players would never consider playing the recommended action with some probability. For $\epsilon$-sCCE, we will see that this requirement can be relaxed.

\subsection{sNE}

To install an sNE, we can always use the utility function for which the desired action's utility is the maximum possible, $B$, and all other utilities are the minimum possible, $-B$. Given the reward bounds, the largest gap that can be enforced is $2B$. Thus, we can only enforce an $\epsilon$ gap if $\epsilon < 2B$.

\begin{proposition}\label{prop: sne-strong}
    Given any pure strategy $a^* \in \mA$, dominance threshold $\epsilon$, and reward bound $B$, there exists a utility function $u \in [-B,B]^{A}$ for which $a^*$ is an $\epsilon$-sNE, for the class of deviations that never play $a^*$ a.s., for $(\mA, u)$ if and only if $\epsilon < 2B$ and deviations strategies may not play $a^*$ a.s.
\end{proposition}

\subsection{sCE}

We next extend these ideas to install $\epsilon$-sCE. Importantly, the proof of \cref{thm: sce} shows a recommended action $j$ beats out a deviation action $k$ by exactly $\norm{\sigma_{ij}}_2(1 - \cos(\theta_{ijk}))$. Differing conditionals ensures this quantity is strictly bigger than $0$, but cannot guarantee it is larger than $\epsilon$. However, if we scale every utility by $\alpha \defeq \frac{\epsilon}{\gamma_i}$ where $\gamma_i \defeq \min_{j \neq k} \norm{\sigma_{ij}}_2(1-\cos(\theta_{ijk}))$, then the dominance gap correspondingly scales up to $\alpha (1 - \gamma_i) = \epsilon$. On the downside, this could push utilities above the reward bound $B$. Consequently, this approach only works when $\frac{\epsilon}{\gamma_i} \leq B$ for all $i$.

\begin{proposition}\label{prop: sce-strong}
    Given any mixed strategy $\sigma \in \Delta(\mA)$, dominance threshold $\epsilon$, and reward bound $B$, there exists a utility function $u \in [-B,B]^{A}$ for which $\sigma$ is an $\epsilon$-sCE, under the class of deviations that never play the recommend action a.s., for the game defined by $u$ if $\epsilon \leq B\gamma^{CE}$, where $\gamma^{CE} \defeq \min_{i,j,k}\norm{\sigma_{ij}}_2(1-\cos(\theta_{ijk}))$.
\end{proposition}

\subsection{sCCE}

For $\epsilon$-sCCE, the proof of \cref{thm: scce} shows that the deviation difference gap for player $i$ for deviation $m$ is lower bounded by $\sum_{j} p_{ij}\norm{\sigma_{ij}}_2(1-\cos(\theta_{ijm}))$. Again, we can scale all utilities by some $\alpha$ to ensure this lower bound exceeds $\epsilon$. Here, we can handle true stochastic deviations so long as the pair-wise condition for standard sCCE holds. This is because if a player ever places a large mass on some pure action, the mixing in $\sigma$ can be used to get more utility from the other supported action. 

\begin{proposition}\label{prop: scce-strong}
    Given any mixed strategy $\sigma \in \Delta(\mA)$, dominance threshold $\epsilon$, and reward bound $B$, there exists a utility function $u \in [-B,B]^{A}$ for which $\sigma$ is an $\epsilon$-sCE for the game defined by $u$ if the pairs condition from \cref{thm: scce} holds and $\epsilon \leq B \gamma^{CCE}$, where $\gamma^{CCE} \defeq \min_{i, m} \sum_{j} \norm{\sigma_{ij}}_2(1-\cos(\theta_{ijm}))$.
\end{proposition}

\subsection{sMPCCE}

We next extend these ideas to enforce $\epsilon$-strictness in Markov games. We can essentially use the same results as for normal-form games. However, due to error in installation over time, the slack must now be scaled by a factor of $1/H$ to ensure strictness holds at each stage.

\begin{proposition}\label{prop: ssmpe}
    Given any policy $\pi \in \Pi$, there exists a reward function $r$ for which $\pi$ is an $\epsilon$-strict Markov-perfect equilibrium (NE/CE/CCE) for $G[r]$ if the conditions for each stage game is satisfied and $\epsilon \leq LB/H$ where LB is the respective bound for the desired solution concept for normal-form games.
\end{proposition}

\section{Conclusion}\label{sec: conclusion}

In this work, we resolved the open question of installability for strict solution concepts. Namely, we derived exact characterizations of what behaviors are possible for normal-form games. We then extended these characterizations to sufficient conditions for $\epsilon$-strict installation and strict installation in Markov games. Our characterizations immediately translate to polynomial time algorithms for verifying a policy's installability and producing simple feasible solutions. We then further these results by deriving Linear Programming based algorithms for cost minimizing reward functions in polynomial time. Our work demonstrates that the mechanism design literature does not have to settle for weak, unpredictable solutions. Strong solutions that ensure game outcome predictability are possible for many desired behavior profiles.

\bibliographystyle{abbrvnat}
\bibliography{refs}

\begin{thebibliography}{19}
\providecommand{\natexlab}[1]{#1}
\providecommand{\url}[1]{\texttt{#1}}
\expandafter\ifx\csname urlstyle\endcsname\relax
  \providecommand{\doi}[1]{doi: #1}\else
  \providecommand{\doi}{doi: \begingroup \urlstyle{rm}\Url}\fi

\bibitem[Anderson et~al.(2010)Anderson, Shoham, and
  Altman]{anderson2010internal}
A.~Anderson, Y.~Shoham, and A.~Altman.
\newblock Internal implementation.
\newblock In \emph{Proceedings of the 9th International Conference on
  Autonomous Agents and Multiagent Systems: volume 1-Volume 1}, pages 191--198.
  Citeseer, 2010.

\bibitem[Banihashem et~al.(2022)Banihashem, Singla, Gan, and
  Radanovic]{banihashem2022admissible}
K.~Banihashem, A.~Singla, J.~Gan, and G.~Radanovic.
\newblock Admissible policy teaching through reward design.
\newblock \emph{arXiv preprint arXiv:2201.02185}, 2022.

\bibitem[Huang and Zhu(2019)]{huang2019deceptive}
Y.~Huang and Q.~Zhu.
\newblock Deceptive reinforcement learning under adversarial manipulations on
  cost signals.
\newblock In \emph{International Conference on Decision and Game Theory for
  Security}, pages 217--237. Springer, 2019.

\bibitem[Ma et~al.(2019)Ma, Zhang, Sun, and Zhu]{ma2019policy}
Y.~Ma, X.~Zhang, W.~Sun, and J.~Zhu.
\newblock Policy poisoning in batch reinforcement learning and control.
\newblock \emph{Advances in Neural Information Processing Systems},
  32:\penalty0 14570--14580, 2019.

\bibitem[Ma et~al.(2021)Ma, Wu, and Zhu]{ma2021game}
Y.~Ma, Y.~Wu, and X.~Zhu.
\newblock Game redesign in no-regret game playing.
\newblock \emph{arXiv preprint arXiv:2110.11763}, 2021.

\bibitem[Monderer and Tennenholtz(2003)]{monderer2003k}
D.~Monderer and M.~Tennenholtz.
\newblock k-implementation.
\newblock In \emph{Proceedings of the 4th ACM conference on Electronic
  Commerce}, pages 19--28, 2003.

\bibitem[Osborne and Rubinstein(1994)]{osborne1994course}
M.~Osborne and A.~Rubinstein.
\newblock \emph{A Course in Game Theory}.
\newblock A Course in Game Theory. MIT Press, 1994.
\newblock ISBN 9780262650403.

\bibitem[Rakhsha et~al.(2020)Rakhsha, Radanovic, Devidze, Zhu, and
  Singla]{rakhsha2020policy}
A.~Rakhsha, G.~Radanovic, R.~Devidze, X.~Zhu, and A.~Singla.
\newblock Policy teaching via environment poisoning: Training-time adversarial
  attacks against reinforcement learning.
\newblock In \emph{International Conference on Machine Learning}, pages
  7974--7984. PMLR, 2020.

\bibitem[Rakhsha et~al.(2021{\natexlab{a}})Rakhsha, Radanovic, Devidze, Zhu,
  and Singla]{rakhsha2021policy}
A.~Rakhsha, G.~Radanovic, R.~Devidze, X.~Zhu, and A.~Singla.
\newblock Policy teaching in reinforcement learning via environment poisoning
  attacks.
\newblock \emph{Journal of Machine Learning Research}, 22\penalty0
  (210):\penalty0 1--45, 2021{\natexlab{a}}.

\bibitem[Rakhsha et~al.(2021{\natexlab{b}})Rakhsha, Zhang, Zhu, and
  Singla]{rakhsha2021reward}
A.~Rakhsha, X.~Zhang, X.~Zhu, and A.~Singla.
\newblock Reward poisoning in reinforcement learning: Attacks against unknown
  learners in unknown environments.
\newblock \emph{arXiv preprint arXiv:2102.08492}, 2021{\natexlab{b}}.

\bibitem[Rangi et~al.(2022)Rangi, Xu, Tran-Thanh, and
  Franceschetti]{rangiunderstanding}
A.~Rangi, H.~Xu, L.~Tran-Thanh, and M.~Franceschetti.
\newblock Understanding the limits of poisoning attacks in episodic
  reinforcement learning.
\newblock In L.~D. Raedt, editor, \emph{Proceedings of the Thirty-First
  International Joint Conference on Artificial Intelligence, {IJCAI-22}}, pages
  3394--3400. International Joint Conferences on Artificial Intelligence
  Organization, 7 2022.
\newblock \doi{10.24963/ijcai.2022/471}.
\newblock Main Track.

\bibitem[Roughgarden(2010)]{roughgarden2010algorithmic}
T.~Roughgarden.
\newblock Algorithmic game theory.
\newblock \emph{Communications of the ACM}, 53\penalty0 (7):\penalty0 78--86,
  2010.

\bibitem[Tadelis(2013)]{tadelis2013game}
S.~Tadelis.
\newblock \emph{Game Theory: An Introduction}.
\newblock Princeton University Press, 2013.

\bibitem[Wu et~al.(2023{\natexlab{a}})Wu, McMahan, Chen, Chen, Zhu, and
  Xie]{wu2023minimally}
Y.~Wu, J.~McMahan, Y.~Chen, Y.~Chen, X.~Zhu, and Q.~Xie.
\newblock Minimally modifying a markov game to achieve any nash equilibrium and
  value.
\newblock \emph{arXiv preprint arXiv:2311.00582}, 2023{\natexlab{a}}.

\bibitem[Wu et~al.(2023{\natexlab{b}})Wu, McMahan, Zhu, and Xie]{wu2023reward}
Y.~Wu, J.~McMahan, X.~Zhu, and Q.~Xie.
\newblock Reward poisoning attacks on offline multi-agent reinforcement
  learning.
\newblock In \emph{Proceedings of the aaai conference on artificial
  intelligence}, volume~37, pages 10426--10434, 2023{\natexlab{b}}.

\bibitem[Wu et~al.(2024)Wu, McMahan, Zhu, and Xie]{wu2024data}
Y.~Wu, J.~McMahan, X.~Zhu, and Q.~Xie.
\newblock Data poisoning to fake a nash equilibria for markov games.
\newblock In \emph{Proceedings of the AAAI Conference on Artificial
  Intelligence}, volume~38, pages 15979--15987, 2024.

\bibitem[Zhang and Parkes(2008)]{zhang2008value}
H.~Zhang and D.~C. Parkes.
\newblock Value-based policy teaching with active indirect elicitation.
\newblock In \emph{AAAI}, volume~8, pages 208--214, 2008.

\bibitem[Zhang et~al.(2009)Zhang, Parkes, and Chen]{zhang2009policy}
H.~Zhang, D.~C. Parkes, and Y.~Chen.
\newblock Policy teaching through reward function learning.
\newblock In \emph{Proceedings of the 10th ACM conference on Electronic
  commerce}, pages 295--304, 2009.

\bibitem[Zhang et~al.(2020)Zhang, Ma, Singla, and Zhu]{zhang2020adaptive}
X.~Zhang, Y.~Ma, A.~Singla, and X.~Zhu.
\newblock Adaptive reward-poisoning attacks against reinforcement learning.
\newblock In \emph{International Conference on Machine Learning}, pages
  11225--11234. PMLR, 2020.

\end{thebibliography}

\appendix

\section{Proof of \texorpdfstring{\cref{prop: ssmpe}}{prop: smpe}}

\begin{proof}
    Fix any player $i \in [n]$. We proceed by induction on $h \in [H+1]$. 
    
    \paragraph{Base Case.} For the base case, we consider $h = H+1$. In this case, there is no reward function, so the claim vacuously holds.

    \paragraph{Inductive Step.} For the inductive step, we consider any $h \leq H$. By the policy evaluation equations, we know that,
    \begin{equation}
        V_{i,h}^{\pi}(s) = \sum_{a \in \mA} \pi_h(a \mid s) \brak{r_h(s,a) + \sum_{s'} P_h(s' \mid s,a)V^{\pi}_{i,h+1}(s')}.
    \end{equation}
    By the inductive hypothesis, we know that $\abs{V^{\pi}_{i,h+1}(s')} \leq B/2$. We define the reward at the current stage by,
    \begin{equation}
        r_h(s,a) \defeq \frac{B}{2}\pi_{ia_{i}}^{hs}(a_{-i}) - \sum_{s'} P_h(s' \mid s,a)V^{\pi}_{i,h+1}(s').
    \end{equation}
    We can use the future value bound and the fact that $\pi_{ia_{i}}^{hs}(a_{-i})$ is a distribution to bound the current rewards as follows:
    \begin{equation}
        \abs{r_h(s,a)} = \abs{\frac{B}{2}\pi_{ia_{i}}^{hs}(a_{-i}) - \sum_{s'} P_h(s' \mid s,a)V^{\pi}_{i,h+1}(s')} \leq \frac{B}{2} + \frac{B}{2} = B.
    \end{equation}
    Thus, the reward satisfies the reward bound. Moreover, we have that $V_{i,h}^{\pi}(s) = \sum_{a \in \mA} \pi_h(a \mid s)\frac{B}{2}\pi_{ia_{i}}^{hs}(a_{-i})$, which implies that $\abs{V_{i,h}^{\pi}(s)} \leq B/2$.

    Now observe that for any deviation policy $\pi' = (\pi_i', \pi_{-i})$,
    \begin{equation}
        V^{\pi}_{i,h}(s) - V^{\pi'}_{i,h}(s) = \frac{B}{H}\sum_{a \in \mA} \pi_h(a \mid s)\brak{\pi_{ia_{i}}^{hs}(a_{-i}) - \pi_{ia_{i}'}^{hs}(a_{-i})} > 0.
    \end{equation}
    The strict inequality follows from the proof of \cref{thm: scce}. This completes the proof.
    
\end{proof}

\begin{algorithm}
    \caption{sCE Installability}\label{alg: sce}
    \begin{algorithmic}[1]
        \Require{$(\mA, \sigma)$}
        \For{$i \in [n]$}
            \For{$j \in \mA_i$}
                \For{$k \in \mA_i$}
                    \If{$j \neq k$ and $\sigma_{ij} = \sigma_{ik} \neq 0$}
                        \State \Return FALSE
                    \EndIf
                \EndFor
            \EndFor
        \EndFor
        \State \Return TRUE
    \end{algorithmic}
\end{algorithm}

\begin{algorithm}
    \caption{sCCE Installability}\label{alg: scce}
    \begin{algorithmic}[1]
        \Require{$(\mA, \sigma)$}
        \For{$i \in [n]$}
            \State $k \gets \argmin\{j \in \mA_i \mid \sigma_{ij} \neq 0\}$
            \State $\textit{differ} \gets$ FALSE
            \For{$j \in \mA_i$}
                    \If{$0 \neq \sigma_{ij} \neq \sigma_{ik}$}
                        \State $\textit{differ} \gets$ TRUE
                    \EndIf
            \EndFor
            \If{not $\textit{differ}$ and $1 \not \in \sigma_{ik}$}
                \State \Return FALSE
            \EndIf
        \EndFor
        \State \Return TRUE
    \end{algorithmic}
\end{algorithm}

\end{document}